# Electromechanical Brillouin scattering in integrated optomechanical waveguides

QIYU LIU,[1,†] HUAN LI,[1,†] AND MO LI[1,2,*]

[1]Department of Electrical and Computer Engineering, University of Washington, Seattle, WA 98195, USA
[2]Department of Physics, University of Washington, Seattle, WA 98195, USA
*moli96@uw.edu

**Abstract:** In the well-known stimulated Brillouin scattering (SBS) process, spontaneous acoustic phonons in materials are stimulated by laser light and scatter the latter into a Stokes sideband. SBS becomes more pronounced in optical fibers and has been harnessed to amplify optical signals and even achieve lasing. Exploitation of SBS has recently surged on integrated photonics platforms as simultaneous confinement of photons and phonons in waveguides leads to drastically enhanced interaction. Instead of being optically stimulated, coherent phonons can also be electromechanically excited with very high efficiency as has been exploited in radiofrequency acoustic filters. Here, we demonstrate electromechanically excited Brillouin scattering in integrated optomechanical waveguides made of piezoelectric material aluminum nitride (AlN). Acoustic phonons of 16 GHz in frequency are excited with nanofabricated electromechanical transducers to scatter counter-propagating photons in the waveguide into a single anti-Stokes sideband. We show that phase-matching conditions of Brillouin scattering can be tuned by varying both the optical wavelength and the acoustic frequency to realize tunable single-sideband modulation. Combining Brillouin scattering photonics with nanoelectromechanical systems, our approach provides an efficient interface between microwave and optical photons that will be important for microwave photonics and potentially quantum transduction.

## 1. Introduction

The scattering of light by the sound wave, namely Brillouin scattering, is of technological importance because this three-wave mixing process provides transduction between the fast-moving photons and the slow-moving phonons [1-7]. The phonons that induce Brillouin scattering can be thermally or electrically excited, or stimulated by laser through optical forces. Stimulated Brillouin scattering (SBS) can occur in optical fibers [8-11], and more recently, in integrated photonic waveguides [12-14] and optical cavities [15-18], in which the optical wave and the acoustic wave can co-propagate. With a strong pump light, the acoustic mode, in fiber or waveguide, that satisfies both energy and momentum conservation (i.e. phase-matching conditions) will be stimulated and scatter the light into a Stokes sideband that has a frequency lower than the pump exactly by the mechanical frequency. Because the mechanical frequency in SBS is typically in multiple gigahertz ranges, SBS has found important application in optical processing of microwave signals [19]. With the advance of integrated photonics, the effort in exploiting SBS recently has shifted from optical fibers to integrated waveguides on material platforms of silicon, silicon nitride, glass and chalcogenide photonics, leading to significant results including Brillouin amplification and lasing, tunable microwave filters, frequency combs and synthesizers [12-14, 20-25].

    Through optically stimulating acoustic waves at microwave frequencies, SBS provides a route of transduction from the optical domain to the microwave domain. On the other hand, acoustic waves can be readily excited electromechanically using piezoelectric materials, which have led to the tremendously successful technology of acoustic RF filters that enabled modern

wireless technology [26, 27]. Using integrated transducers, microwave signals can be converted to acoustic waves and vice versa with very low loss [28-30]. Realizing on-chip Brillouin scattering using electromechanically excited acoustic waves, as opposed to optical stimulation in SBS, thus will provide the complementary and equally important transduction from the microwave to the optical domain. In this work, we demonstrate electromechanical Brillouin scattering in optomechanical waveguides in which photons and acoustic phonons are co- or counter-propagating. Acoustic phonons at a frequency as high as 16 GHz are excited with an electromechanical transducer and coupled into the waveguide. The ultrahigh frequency phonons provide the momentum required by the phase-matching condition to backscatter counter-propagating photons into an anti-Stokes sideband (ASB). This acousto-optic scattering process distinguishes from those in prior work in guided-wave acousto-optics [31-35], in which acoustic wave of much lower frequency is used to deflect light by only a small angle in a 2D waveguide system. We further demonstrate transduction of RF signals to a single sideband of the optical carrier and conversion back to the RF domain at a receiver with preserved phase information. With gigahertz acoustic frequency, which is one or two orders of magnitude higher than that in the conventional acousto-optics devices, our device promises to drastically increase the operation bandwidth for applications in future microwave photonics signal processing. The ultrahigh frequency also enabled the unprecedented backward Brillouin scattering configuration in the OM waveguide, which, when fully optimized, can naturally achieve tunable single-sideband modulation within a compact device footprint.

## 2. Electromechanical Brillouin scattering in an optomechanical waveguide

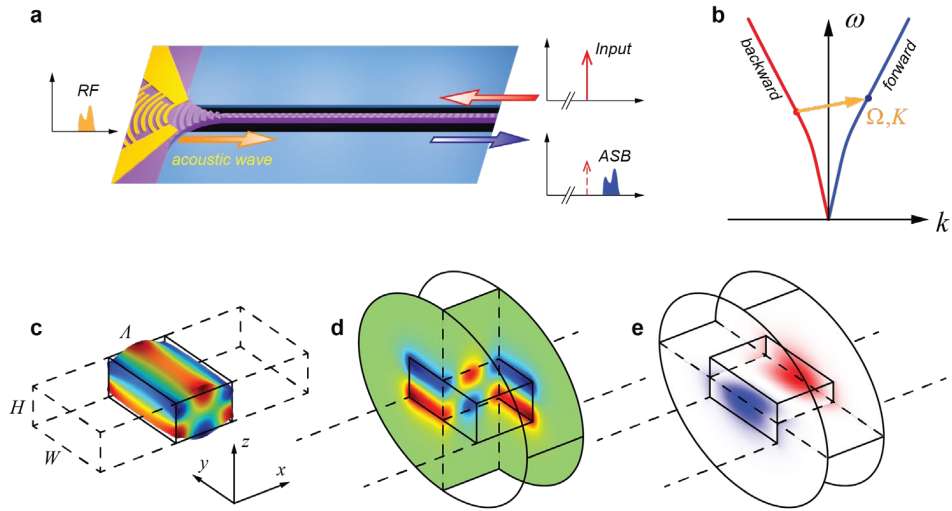

**Fig. 1.** Electromechanical Brillouin scattering in an optomechanical waveguide. a. Schematic illustration of an optomechanical waveguide. The acoustic wave is excited with the IDT and propagates in the forward direction to scatter backward propagating optical carrier into the anti-Stokes sideband (ASB). b. Simplified dispersion diagram of the optical mode in the waveguide. The acoustic mode (orange line) that satisfies the phase-matching conditions scatters the backward propagating optical mode (red line) to the forward mode (blue line). c-e. Finite-element simulation results of the acoustic S0 mode, displayed in displacement (c), and piezoelectric potential (d), and the optical TE0 mode (e).

We build the entire device on the aluminum nitride (AlN) on silicon platform, where both the optical waveguides and the piezoelectric transducers are fabricated on the 330 nm thick AlN layer [36-40]. As illustrated in Fig. 1a, to afford simultaneous confinement and propagation of both photons and phonons, the AlN optomechanical waveguide is suspended from the substrate [41]. In the forward direction, with an interdigital transducer (IDT), RF signal is converted to an acoustic wave, which is focused into and propagates along the waveguide. In the backward direction, an optical carrier is input to the waveguide, counter-propagating with the acoustic wave. When phase-matching conditions are satisfied, efficient Brillouin scattering occurs such that the optical carrier is scattered back by the acoustic wave, generating a single upper sideband—the anti-Stokes sideband (ASB)—at the frequency exactly higher than the optical carrier by the acoustic frequency. The dispersion relation of this three-wave mixing process is illustrated in Fig. 1b. Because the acoustic frequency is much lower than the optical frequency, the wavevector magnitudes of the input and the backscattered optical modes are approximately the same. Thus, phase-matching requires $|q| = 2|\beta|$, where $q$ and $\beta$ are the wavevectors of the acoustic and the optical modes, respectively. Hence, the acoustic wavelength $\Lambda$ needs to be half of the optical wavelength $\lambda = \lambda_0/n_e$ in a waveguide with an effective index of $n_e$, that is, the Bragg condition of reflection. Consider an AlN waveguide with $n_e = 1.5$ in the near-infrared telecom band ($\lambda_0 = 1550$ nm), the required acoustic wavelength will be $\Lambda = 500$ nm and the corresponding frequency ($f = v_p/\Lambda$, where $v_p$ is acoustic phase velocity) will be around 16 GHz, assuming $v_p = 8 \times 10^3$ m/s. Previously, we have shown that, in suspended AlN thin film, acoustic wave of this high frequency can be generated efficiently in the form of symmetric Lamb modes [41].

Finite-element simulation result of one full wavelength of the fundamental symmetric Lamb mode (S0 mode) in the AlN waveguide is shown in Fig. 1c and d. Because the AlN thin film used is polycrystalline with highly aligned c-axis in the out-of-the-plane direction, its properties relevant to the acoustic mode are isotropic in the plane, which is assumed in the simulation. A representative waveguide design is used in the simulation, where the width ($W$), thickness ($H$), the acoustic wavelength ($\Lambda$) and the frequency ($f$) are 800 nm, 330 nm, 500 nm, and 16.248 GHz, respectively. The dominant mechanical motion of the mode is the vertical deformation (extension and compression) (Fig. 1c), which accordingly induces piezoelectric potential (Fig. 1d) with periodically alternating polarity along the waveguide. Fig. 1e shows the transverse electric field component $E_y$ of the fundamental TE (TE0) optical mode with a vacuum optical wavelengths ($\lambda_0$) of 1.5 μm. Because this TE0 mode's wavelength in the waveguide $\lambda = 1.0$ μm ($n_e = 1.5$), it is phase-matched with the acoustic S0 mode, hence expected to be strongly scattered.

## 3. Ultrahigh frequency transducer exciting acoustic wave to an optomechanical waveguide

Fig. 2a and b show scanning electron microscope (SEM) images of a representative device. The devices are fabricated with standard nanofabrication processes (see Methods section for details). In the last step of fabrication, XeF$_2$ etching is used to undercut-etch the silicon to suspend the waveguides and the IDT region. The core component in the device is the straight free-standing AlN optomechanical (OM) waveguide, as shown in Fig. 2a. The dry etching process yields an estimated sidewall angle of 75° such that the waveguide has a trapezoid cross-section. In the study, the width and length of the waveguide are varied in different devices.

At the front end of the OM waveguide, a curved split-finger IDT on the suspended AlN membrane is used to excite acoustic waves from the input electrical RF signal [42, 43]. The acoustic waves subsequently couple into the waveguide through a compact horn structure [44]. At the back of the IDT, an etched-through trench is used to reflect the backward-propagating acoustic wave into the forward direction to improve efficiency. Based on simulation, the acoustic wavelength $\Lambda$ needs to be 500 nm. Accordingly, as shown in the inset of Fig. 2a, the IDT period and the individual finger width are designed to be 1.5 μm and 187.5 nm, respectively, such that the wavelength of the 3rd spatial harmonic the split-finger IDT can efficiently excites is exactly 500 nm[45, 46]. In the study, both the IDT period and RF frequency are varied in a small range around the design values. The IDT fingers consist of a 140 nm thick aluminum bottom layer and a 25 nm thick gold top layer, with the latter passivating and protecting the aluminum layer. At the back end of the OM waveguide (Fig. 2b), a taper structure is used to adiabatically couple between the fundamental TE optical mode of the OM waveguide and that of the suspended ridge waveguide in the peripheral integrated photonic circuit, which consists of waveguides, directional couplers and input/output grating couplers.

The equivalent circuit model of the acoustic transducer is shown in Fig. 2c. The resistor $R_s$ accounts for the total series resistance between the RF probe and the IDT fingers. The shunt resistor $R_l$ and the capacitor $C_e$ account for the effective leakage resistance and electrode capacitance between the IDT fingers, respectively. The electromechanical response of the transducer is modeled with a complex and frequency dependent admittance $Y_a$, following the Butterworth Van Dyke (BVD) model [47, 48]. The electrical power dissipated on $Y_a$ is converted into acoustic power.

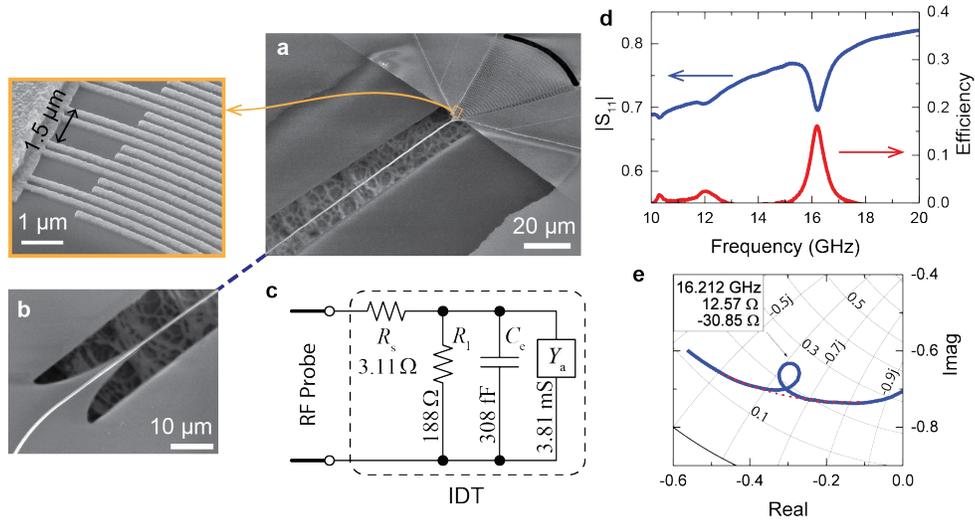

**Fig. 2.** Ultrahigh frequency transducer exciting acoustic wave to an optomechanical waveguide. a. Scanning electron microscope (SEM) image of the device, featuring the suspended optomechanical waveguide and the curved IDT to excite and focus acoustic wave. Inset: Zoom-in view of the IDT, showing the split finger design with an electrode width of 187.5 nm. b. Detailed view of the optical input to the optomechanical waveguide, which uses a taper to convert a rib waveguide to a fully suspended ridge waveguide. c. Equivalent circuit model of the IDT acoustic transducer with component values extracted from the data in (d) and (e). d. Microwave reflectance ($S_{11}$ parameter) measurement of the IDT, showing a strong acoustic mode at 16.2 GHz. e. Vectoral analysis of the IDT plotted on a Smith chart. The on-resonance impedance is 12.57-$j$30.85 Ohm.

The RF reflection coefficient ($S_{11}$), measured with a vector network analyzer (VNA) from a representative device is shown in Figs. 2d and e. Fitting the results with the equivalent circuit model yields the values of the components, as labeled in Fig. 2c, where the value of $Y_a$ is that for the S0 mode resonance. In Fig. 2d, $S_{11}$ shows a major resonance for the S0 mode at 16.2 GHz, as well as two weak resonances for other acoustic modes at 10.3 GHz and 12.0 GHz. The electromechanical conversion efficiency is calculated from the circuit model and shown in Fig.

2d, which peaks at the S0 resonance with a maximum efficiency of 17%. The split-finger IDT also excites the fundamental mode at 7.65 GHz but with a much lower efficiency of 2.3% because of power impedance matching. In Fig. 2e, $S_{11}$ is plotted on a Smith chart, which shows the S0 mode resonance as a circle on a capacitive background dominated by $C_e$. The result shows that the on-resonance impedance is 12.57-$j$30.85 Ω, thus is not impedance-matched with the 50 Ω input transmission line. Achieving impedance-matching by using a shunt inductor can significantly improve the coupling efficiency with the signal source. For example, modern acoustic filters can achieve an insertion loss of 1 dB, which corresponds to an electrotechnical conversion efficiency of 90% [28, 29].

## 4. Experimental demonstration of electromechanical Brillouin scattering

To characterize the Brillouin scattering process, we used the measurement scheme illustrated in Fig. 3a. The RF signal from the VNA is input to the IDT to excite acoustic wave propagating in the forward direction. An optical carrier from a tunable laser is coupled to the chip via a grating coupler and to the OM waveguide in the backward direction. Any scattered light in the forward direction couples to a directional coupler and outputs to an optical spectrum analyzer (OSA) with a resolution bandwidth of 20 pm. Fig. 3b shows the output spectra measured from a representative device with a 5 μm wide, 500 μm long OM waveguide. The input optical carrier power in the waveguide, the vacuum wavelength, and the RF frequency were all fixed at −4.4 dBm, 1510 nm, and 16.4 GHz, respectively. When the RF power was turned on and increased from 7.9 dBm to 20.9 dBm, a sideband peak appeared on the spectra with increasing magnitude and a frequency exactly 16.4 GHz higher than the optical carrier. This is clearly the anti-Stokes sideband (ASB) due to Brillouin scattering between the counter-propagating optical and acoustic waves happening in the OM waveguide. The absence of the Stokes sideband indicates that the acousto-optic scattering processes occurred mainly in the OM waveguide, instead of elsewhere such as in the IDT region, where the bidirectionally-propagating acoustic waves would have generated both the Stokes and the anti-Stokes sidebands. The center peak in the spectra, which remain unchanged with the RF power, is the reflected optical carrier due to Rayleigh scattering occurring in the OM waveguide and at the IDT. Theoretically, the ASB power should increase linearly with the acoustic power, which is proportional to the input RF power. The power of the ASB peak is extracted from the spectra and plotted versus the corresponding RF power in the inset of Fig. 3b. The measured results show clear linear dependence when the RF power is below 16 dBm. Above 16 dBm, however, a super-linear dependence of the ASB power on the RF power is observed. We tentatively attribute this super-linearity to the heating effect at high RF power levels which may cause changes in materials' properties.

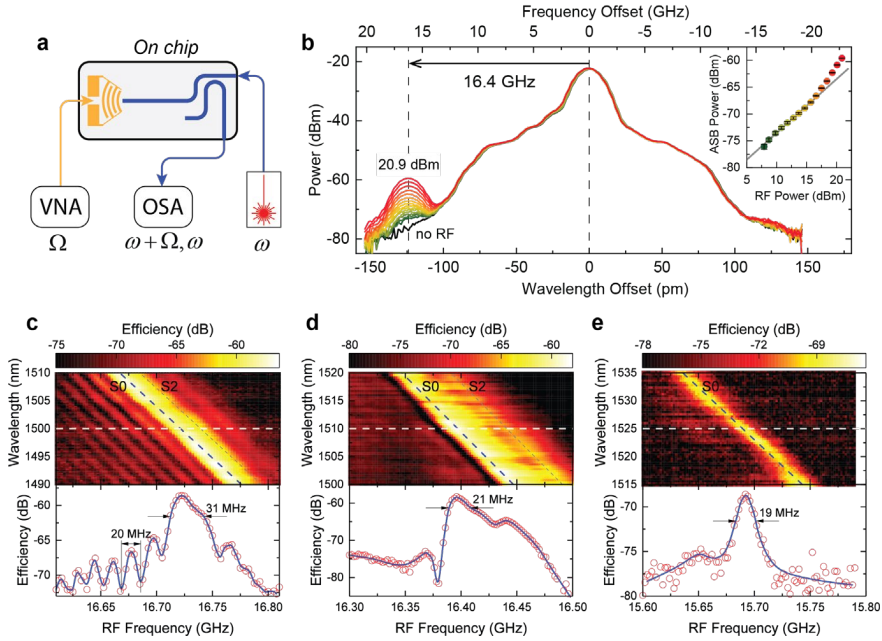

**Fig. 3.** Experimental demonstration of electromechanical Brillouin scattering. a. Diagram of the measurement setup. b. Spectra of the Brillouin scattered optical signal, showing a single anti-Stokes sideband (ASB) offset from the carrier by 16.4 GHz. Inset: the ASB power versus input RF power. At relatively high RF power, the dependence becomes super-linear. c-e. 2D plot of the ASB power when both the optical wavelength and the RF frequency are scanned, measured from devices with waveguide dimension of 5 μm × 100 μm (c), 5 μm × 500 μm (d), and 0.8 μm × 600 μm (e), respectively. Lower panels show the cross-sectional plot along the white dashed line in the upper panel. Solid blue lines are guides of for the eye. The ASB power peaks along the phase-matching curve marked with the blue dashed line.

To systematically investigate the Brillouin scattering process in the OM waveguide, we measured the ASB power while scanning both the input optical wavelength and the RF frequency with their power fixed. At each combination of the scanned wavelength and frequency, the Brillouin scattering efficiency ($\eta=P_{ASB}/P_0$, where $P_{ASB}$ is the ASB power and $P_0$ is the optical carrier power) is calculated. The results measured from three devices with OM waveguide dimension of 5 μm×100 μm, 5 μm ×500 μm, and 0.8 μm ×600 μm, respectively, are presented as color-coded 2D plots in Fig. 3c, d, and e. Cross-sectional plots at fixed optical wavelengths are also shown. In all three cases, the efficiency reaches a maximum when the wavelength and frequency satisfy the exact phase-matching conditions ($\lambda_0 f=2n_e v_p$), which is a curve that connects the peak points of each cross-sectional curve in the 2D plots. The 2D plots are approximately invariant along this major phase-matching line, whose slope is $\gamma=-0.20$ nm/MHz, which is consistent with the simulated optical TE0 mode and acoustic S0 mode dispersion curves. From the plots, the acoustic phase and group velocities of the S0 mode around 16 GHz in all three devices can be extracted to be similar values of $v_p=8.5\times10^3$ m/s and $v_g=3.0\times10^3$ m/s, respectively.

In addition to the common phase-matching line described above, the 2D plot for each waveguide dimension also shows different unique characteristics. For the 5 μm×100 μm waveguide, the 2D plot exhibits a decaying periodic fringe pattern away from the major phase-matching line, resembling a $\mathrm{sinc}^2(x)$ function. In contrast, for the 5×500 and 0.8×600 waveguides, the 2D plots exhibit a single peak pattern along the phase-matching curve, resembling the Lorentzian function. Such observations agree excellently with the coupled-mode theory (CMT) [49-51] of the three-wave mixing process of Brillouin scattering in OM waveguides (see Supplementary Information for detailed CMT analysis). CMT gives that the pattern of the Brillouin scattering efficiency as a function of the optical wavelength and the RF frequency is determined by the acoustic propagation length ($L_b$) and the effective waveguide length ($L$). In the OM waveguide, the acoustic wave is excited at one end and propagates toward the other end with an exponential amplitude decay constant $\alpha = 1/(2L_b)$. Compared with the acoustic propagation loss, the optical propagation loss is negligibly small. In the short waveguide limit, where $L \ll L_b$, where the acoustic wave does not decay significantly when it reaches the other end of the waveguide, the Brillouin scattering efficiency at a fixed wavelength or RF frequency follows a sinc-squared function of $\text{sinc}^2(\kappa L/2)$ and is independent on $L_b$. The interval between the minima of the $\text{sinc}^2$ function at any fixed wavelength in the 2D plots equals $v_g/L$, the reciprocal of the transit time ($L/v_g$) of the acoustic wave through the waveguide. For the 5×100 waveguide, the short waveguide limit applies, and the 20 MHz interval between two adjacent minima in the cross-sectional curve indicates the effective waveguide length to be ~150 µm, which includes the taper region. It also indicates that the acoustic propagation length $L_b$ has a lower bound of ~150 µm. In the long waveguide limit of $L \gg L_b$, where the acoustic wave decays to near zero power before it reaches the other end of the waveguide, the Brillouin scattering efficiency approximates a Lorentzian function and is independent of $L$. The full-width half-maximum (FWHM) linewidth of the Lorentzian at any fixed wavelength is given by $\Gamma = v_g/(2\pi L_b)$ (See SI for derivation). For the 5×500 and 0.8×600 waveguides, the long waveguide limit applies. The 20 MHz linewidths in Fig. 3c and d, taking $v_g = 3.0×10^3$ m/s, indicates a surprisingly low $L_b$ value of 25 µm, much less than that measured in the short device. We attribute the drastic difference in acoustic loss between the short and long OM waveguides to the variation of AlN film quality in different batches and the unstable etching process in our facility.

Furthermore, in the 5×100 and 5×500 waveguides, other than the dominant S0 mode, another acoustic mode is also observed on the high-frequency side of the phase-matching curves. Although this secondary acoustic mode does not significantly contribute to the Brillouin scattering efficiency, its presence distorts the patterns of the 2D plots, rendering them asymmetric with respect to the phase-matching curves. Such asymmetry is not observed in the 0.8×600 waveguide, indicating that only the S0 mode can propagate in the waveguide.

The highest Brillouin scattering efficiency achieved for the 5×100, 5×500 and 0.8×600 waveguides are −56 dB, −58 dB, and −67 dB, respectively. The above measurement results are drastically lower than the near-unity efficiency achievable with only 1 mW of acoustic power as predicted from our theoretical calculations (see SI). We attribute the drastic difference largely to the unexpectedly high acoustic loss in the long waveguides and the low acoustic coupling efficiency between the IDT and the OM waveguide such that the acoustic power in the waveguide is much less than 1 mW. All of these can be improved with optimized acoustic coupling scheme, fabrication process and material quality. Nevertheless, the results represent the first demonstration of backward Brillouin scattering by electromechanically excited phonons in an integrated waveguide.

## 5. RF photonic link using electromechanical Brillouin scattering

The Brillouin scattering in the OM waveguides naturally provides acousto-optic single-sideband (SSB) modulation that is needed for RF photonic communication but otherwise can only be realized with a sophisticated electro-optic modulator [52-55]. To demonstrate this, we employed both the homodyne and heterodyne interferometry schemes asshown in Fig. 4a.. On the transmitter side, the measurement scheme is the same as that in Fig. 3a, except that the OSA is now replaced by a meters-long fiber link, which transmits both the Brillouin-scattered ASB and the Rayleigh-scattered optical carrier to the receiver side. The entire OM waveguide device now serves as an SSB modulator with a tunable transmission band for the communication link. In the homodyne scheme, on the receiver side, the Rayleigh-scattered optical carrier, being almost constant throughout the measurement, conveniently serves as the local oscillator (LO).

Both the ASB and the carrier transmitted through the fiber link are amplified with an erbium-doped fiber amplifier (EDFA) and routed to a high-speed photodetector (PD). The beating of the ASB and the carrier at the PD down-converts the signal to the RF frequency $\Omega$. The complex amplitude of the beating RF signal is measured with the VNA. Fig. 4b shows the normalized power from the 0.8×600 waveguide measured with four different carrier wavelengths. Expectedly, the efficiency peak moves as the carrier wavelength changes, consistent with the phase-matching line with a slope of −0.20 nm/MHz as in Fig. 3. The efficiency peak for 1530 nm is consistent with the cross-sectional curve at the same wavelength in Fig. 3e. Also included in Fig. 4b is the normalized complex amplitude of the ASB for 1530 nm in a polar chart, showing both the magnitude and the phase. The shaded blue lines in both panels of Fig. 4b show the curve fitting results based on the coupled-mode theory, which approximate a Lorentzian peak and agree excellently with the measurement results.

For heterodyne interferometry, on the receiver side, the LO is derived by frequency-shifting the laser source by an offset frequency $\Delta\omega$. The frequency shift is achieved by an acousto-optic frequency shifter (AOFS) operating at $\Delta\omega/(2\pi)$ = 100 MHz. The ASB transmitted through the fiber link is combined with the LO through a 1:1 fiber directional coupler, and routed to a high-speed photodetector (PD), which generates the beating signal at the RF frequency $\Omega-\Delta\omega$. It is solely for the convenience to derive the LO from the transmitter laser source. In real-world applications, the LO will be another independent laser at the receiver site, whose frequency is most likely different from that of the transmitter laser. Because the ASB and the LO propagate through different fibers, their relative phase is scrambled and thus incoherent due to the random refractive index fluctuations in the fibers. In real-world applications, this will also be the case because the long-haul fiber link will be susceptible to disturbance, and the LO will be from an independent laser source. In contrast, the homodyne interferometry does not suffer from this problem because the ASB and the carrier (LO) always propagate through the same optical path. With scrambled phase, only the power of the beating signal at frequency $\Omega-\Delta\omega$ is measured by the VNA. Fig. 4c shows the measurement results from the 5×500 waveguide for seven different carrier wavelengths, consistent with those in Fig. 3d. The horizontal axis is the frequency $\Omega/2\pi$, at which the IDT is driven. In both the homodyne and heterodyne schemes, the transmission band, which is about 20 MHz to 40 MHz wide, can be tuned continuously across a frequency span of ~200 MHz and >300 MHz, respectively, by changing the input optical carrier wavelength.

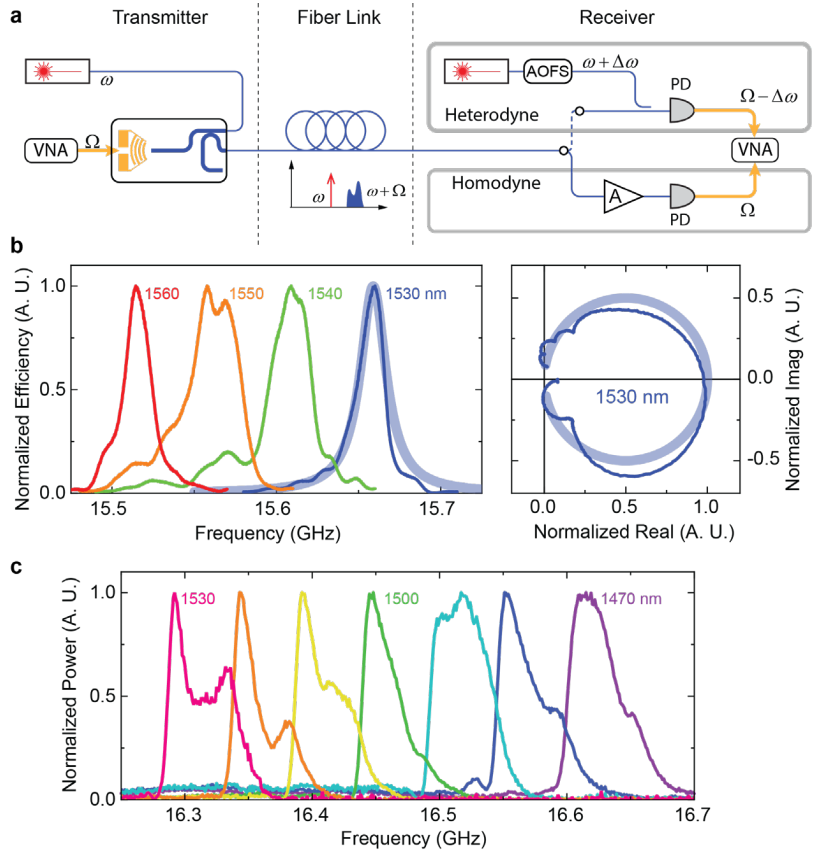

**Fig. 4.** RF photonic link using electromechanical Brillouin scattering. a. Schematics for transmitting RF signals using homodyne and heterodyne demodulation schemes. b. Measurement results when the homodyne scheme is used. Left panel: The transmission band is tunable over a range of ~200 MHz along the phase-matching curve (Fig. 3d) when the optical carrier wavelength is tuned. Right panel: The phase of the transmitted signal is preserved as the received signal shows a full phase tuning between ±π/2 within the transmission band. c. Measurement results when the heterodyne scheme is used. The heterodyne scheme is immune to phase scrambling during the transmission. The transmission band is also tunable over a range of >300 MHz by changing the optical carrier wavelength.

## 6. Conclusion

In conclusion, we have demonstrated, for the first time, backward Brillouin scattering between electromechanically excited acoustic wave and optical wave in an integrated waveguide. In this 1D optomechanical system, the phase-matching conditions of this nonlinear three-wave mixing process are reached with an ultrahigh acoustic frequency of 16 GHz. At this microwave frequency, the acoustic wave has a shorter wavelength than light in the same medium yet travels more than four orders of magnitudes slower. Thus, the acoustic wave has a unique position to bridge the frequency-wavelength gap between microwave and light in integrated photonic devices via this unprecedented acousto-optic scattering process. In the future, the acoustic coupling efficiency into the narrow OM waveguides can be greatly improved by further optimization of acoustic horn and IDT designs. Reduction of the acoustic loss can be achieved by optimizing the dry etching process for the OM waveguides, as well as exploring new thin film materials such as lithium niobate and single-crystal AlN. Through the reduction of acoustic propagation loss and the enhancement of acoustic coupling, the Brillouin scattering efficiency could approach unity so that the demonstrated Brillouin optomechanical system will find important applications in microwave photonics and quantum transduction [56, 57].

**Appendix A: Methods**

*A.1 Device fabrication*

Devices are fabricated on 330 nm thick polycrystalline aluminum nitride film with highly aligned out-of-plane c-axis orientation, deposited by reactive sputtering on silicon wafers. Photonic waveguides and grating couplers are patterned with electron beam lithography (Vistec 5000+ system) using HSQ as the resist. The photonic layer is then etched into the AlN layer by 200 nm using $Cl_2$/$BCl_3$/Ar based ICP/RIE plasma etching (Oxford Plasmalab 100). Releasing windows and the optomechanical waveguide is subsequently patterned with electron beam lithography using ZEP520A as the resist. Another step of ICP/RIE plasma etching is then applied to etch through the AlN layer to open releasing windows. A third ebeam writing step is used to pattern the interdigital transducers (IDTs) using 450 nm PMMA as the resist. A film of 140 nm aluminum and 25 nm gold is deposited with an electron beam evaporator and the IDTs are formed with the liftoff methods. The contact pads are thickened to reduce contact resistance by depositing 600 nm thick Al/Cu/Al/Au layer. Then the trench acoustic reflectors are patterned with double-spun ZEP520A and etched with another step of ICP/RIE etching process. Finally, the device layer is released from the silicon substrate by undercut etching using a $XeF_2$ etcher (SPS Xactix e2).

**Funding**

National Science Foundation (NSF) (EFMA-1641109).

**Acknowledgments**

Parts of this work were carried out in the Minnesota Nano Center, which is supported by the National Science Foundation through the National Nano Coordinated Infrastructure Network (NNCI) under Award Number ECCS-1542202, and the Washington Nanofabrication Facility / Molecular Analysis Facility, a National Nanotechnology Coordinated Infrastructure (NNCI) site at the University of Washington, which is supported in part by funds from the National Science Foundation (awards NNCI-1542101, 1337840 and 0335765), the National Institutes of Health, the Molecular Engineering & Sciences Institute, the Clean Energy Institute, the Washington Research Foundation, the M. J. Murdock Charitable Trust, Altatech, ClassOne Technology, GCE Market, Google and SPTS.

See Supplement 1 for supporting content.

[†]These authors contributed equally to this work

**References**

1. L. Brillouin, "Diffusion de la lumière et des rayons X par un corps transparent homogène," Annales de Physique **9**, 88-122 (1922).

2. R. Chiao, C. Townes, and B. Stoicheff, "Stimulated Brillouin Scattering and Coherent Generation of Intense Hypersonic Waves," Physical Review Letters **12**, 592-595 (1964).

3. C. Quate, C. Wilkinson, and D. Winslow, "Interaction of light and microwave sound," Proceedings of the IEEE **53**, 1604-1623 (1965).

4. R. Adler, "Interaction between light and sound," IEEE spectrum **4**, 42-54 (1967).

5. D. Heiman, D. Hamilton, and R. Hellwarth, "Brillouin scattering measurements on optical glasses," Physical Review B **19**, 6583 (1979).

6. R. W. Boyd, *Nonlinear optics* (Academic Press, 2008).

7. E. Garmire, "Perspectives on stimulated Brillouin scattering," New Journal of Physics **19**, 011003 (2017).

8. E. P. Ippen, and R. H. Stolen, "Stimulated Brillouin scattering in optical fibers," Applied Physics Letters **21**, 539-541 (1972).

9. R. Shelby, M. Levenson, and P. Bayer, "Guided acoustic-wave Brillouin scattering," Physical Review B **31**, 5244 (1985).

10. M. S. Kang, A. Brenn, and P. S. J. Russell, "All-optical control of gigahertz acoustic resonances by forward stimulated interpolarization scattering in a photonic crystal fiber," Physical review letters **105**, 153901 (2010).

11. J.-C. Beugnot, S. Lebrun, G. Pauliat, H. Maillotte, V. Laude, and T. Sylvestre, "Brillouin light scattering from surface acoustic waves in a subwavelength-diameter optical fibre," Nature communications **5**, 5242 (2014).

12. R. Pant, C. G. Poulton, D. Y. Choi, H. Mcfarlane, S. Hile, E. Li, L. Thevenaz, B. Luther-Davies, S. J. Madden, and B. J. Eggleton, "On-chip stimulated Brillouin scattering," Opt Express **19**, 8285-8290 (2011).

13. P. T. Rakich, C. Reinke, R. Camacho, P. Davids, and Z. Wang, "Giant Enhancement of Stimulated Brillouin Scattering in the Subwavelength Limit," Phys Rev X **2**, 011008 (2012).

14. R. Van Laer, B. Kuyken, D. Van Thourhout, and R. Baets, "Interaction between light and highly confined hypersound in a silicon photonic nanowire," Nature Photonics **9**, 199-203 (2015).

15. T. Carmon, H. Rokhsari, L. Yang, T. J. Kippenberg, and K. J. Vahala, "Temporal behavior of radiation-pressure-induced vibrations of an optical microcavity phonon mode," Physical Review Letters **94**, - (2005).

16. M. Tomes, and T. Carmon, "Photonic micro-electromechanical systems vibrating at X-band (11-GHz) rates," Physical review letters **102**, 113601 (2009).

17. G. Bahl, J. Zehnpfennig, M. Tomes, and T. Carmon, "Stimulated optomechanical excitation of surface acoustic waves in a microdevice," Nat Commun **2**, 403 (2011).

18. G. Bahl, M. Tomes, F. Marquardt, and T. Carmon, "Observation of spontaneous Brillouin cooling," Nat Phys **8**, 203-207 (2012).

19. B. J. Eggleton, C. G. Poulton, and R. Pant, "Inducing and harnessing stimulated Brillouin scattering in photonic integrated circuits," Advances in Optics and Photonics **5**, 536-587 (2013).

20. H. Shin, W. Qiu, R. Jarecki, J. a. Cox, R. H. Olsson, A. Starbuck, Z. Wang, and P. T. Rakich, "Tailorable stimulated Brillouin scattering in nanoscale silicon waveguides," Nature communications **4**, 1944-1944 (2013).

21. J. Li, H. Lee, and K. J. Vahala, "Microwave synthesizer using an on-chip Brillouin oscillator," Nature communications **4**, 2097 (2013).

22. D. Marpaung, B. Morrison, M. Pagani, R. Pant, D.-Y. Choi, B. Luther-Davies, S. J. Madden, and B. J. Eggleton, "Low-power, chip-based stimulated Brillouin scattering microwave photonic filter with ultrahigh selectivity," Optica **2**, 76 (2015).

23. M. Merklein, I. V. Kabakova, T. F. S. Büttner, D.-Y. Choi, B. Luther-Davies, S. J. Madden, and B. J. Eggleton, "Enhancing and inhibiting stimulated Brillouin scattering in photonic integrated circuits," Nature Communications **6**, 6396-6396 (2015).

24. E. A. Kittlaus, H. Shin, and P. T. Rakich, "Large Brillouin amplification in silicon," Nature Photonics **10**, 463 (2016).

25. N. T. Otterstrom, R. O. Behunin, E. A. Kittlaus, Z. Wang, and P. T. Rakich, "A silicon Brillouin laser," Science **360**, 1113-1116 (2018).

26. C. Campbell, *Surface acoustic wave devices for mobile and wireless communications* (Academic Press, 1998).

27. D. P. Morgan, *Surface acoustic wave filters : with applications to electronic communications and signal processing* (Academic Press, 2007).

28. K. Yamanouchi, and Y. Satoh, "Ultra low-insertion-loss surface acoustic wave filters using unidirectional interdigital transducers with grating SAW substrates," Japanese journal of applied physics **44**, 4532 (2005).

29. R. Lu, T. Manzameque, Y. Yang, and S. Gong, "S0-Mode Lithium Niobate Acoustic Delay Lines with 1 dB Insertion Loss," in *2018 IEEE International Ultrasonics Symposium (IUS)*(IEEE2018), pp. 1-9.

30. M. K. Ekström, T. Aref, J. Runeson, J. Björck, I. Boström, and P. Delsing, "Surface acoustic wave unidirectional transducers for quantum applications," Applied Physics Letters **110**, 073105 (2017).

31. C. S. Tsai, "Integrated Acousto-Optic Device Modules and Applications," in *Guided-Wave Acousto-Optics*(Springer, 1990), pp. 273-316.

32. L. Kuhn, P. Heidrich, and E. Lean, "Optical guided wave mode conversion by an acoustic surface wave," Applied Physics Letters **19**, 428-430 (1971).

33. Y. Ohmachi, and J. Noda, "LiNbO3 TE-TM mode converter using collinear acoustooptic interaction," IEEE Journal of Quantum Electronics **13**, 43-46 (1977).

34. A. M. Matteo, C. S. Tsai, and N. Do, "Collinear guided wave to leaky wave acoustooptic interactions in proton-exchanged LiNbO/sub 3/waveguides," IEEE transactions on ultrasonics, ferroelectrics, and frequency control **47**, 16-28 (2000).

35. D. B. Sohn, S. Kim, and G. Bahl, "Time-reversal symmetry breaking with acoustic pumping of nanophotonic circuits," Nature Photonics **12**, 91 (2018).

36. W. H. P. Pernice, C. Xiong, C. Schuck, and H. X. Tang, "High-Q aluminum nitride photonic crystal nanobeam cavities," Applied Physics Letters **100**, 091105 (2012).

37. C. Xiong, W. H. Pernice, and H. X. Tang, "Low-loss, silicon integrated, aluminum nitride photonic circuits and their use for electro-optic signal processing," Nano letters **12**, 3562-3568 (2012).

38. S. A. Tadesse, and M. Li, "Sub-optical wavelength acoustic wave modulation of integrated photonic resonators at microwave frequencies," Nature communications **5** (2014).

39. H. Li, S. A. Tadesse, Q. Liu, and M. Li, "Nanophotonic cavity optomechanics with propagating acoustic waves at frequencies up to 12 GHz," Optica **2**, 826 (2015).

40. L. Fan, C.-L. Zou, M. Poot, R. Cheng, X. Guo, X. Han, and H. X. Tang, "Integrated optomechanical single-photon frequency shifter," Nature Photonics **10**, 766 (2016).

41. S. A. Tadesse, H. Li, Q. Liu, and M. Li, "Acousto-optic modulation of a photonic crystal nanocavity with Lamb waves in microwave K band," Applied Physics Letters **107**, 201113 (2015).

42. M. de Lima Jr, M. Beck, R. Hey, and P. Santos, "Compact Mach-Zehnder acousto-optic modulator," Applied physics letters **89**, 121104 (2006).

43. D. C. Malocha, "Evolution of the SAW transducer for communication systems," in *Ultrasonics Symposium, 2004 IEEE*(IEEE2004), pp. 302-310.

44. M. M. de Lima Jr, and P. V. Santos, "Modulation of photonic structures by surface acoustic waves," Reports on progress in physics **68**, 1639 (2005).

45. H. Engan, "Surface Acoustic Wave Multielectrode Transducers," IEEE Transactions on Sonics and Ultrasonics **22**, 395-401 (1975).

46. M. M. d. L. Jr., F. Alsina, W. Seidel, and P. V. Santos, "Focusing of surface-acoustic-wave fields on (100) GaAs surfaces," Journal of Applied Physics **94**, 7848-7855 (2003).

47. J. D. Larson, P. D. Bradley, S. Wartenberg, and R. C. Ruby, "Modified Butterworth-Van Dyke circuit for FBAR resonators and automated measurement system," in *Ultrasonics Symposium, 2000 IEEE*(IEEE2000), pp. 863-868.

48. I. Camara, B. Croset, L. Largeau, P. Rovillain, L. Thevenard, and J.-Y. Duquesne, "Vector network analyzer measurement of the amplitude of an electrically excited surface acoustic wave and validation by X-ray diffraction," Journal of Applied Physics **121**, 044503 (2017).

49. C. Wolff, M. J. Steel, B. J. Eggleton, and C. G. Poulton, "Stimulated Brillouin scattering in integrated photonic waveguides: Forces, scattering mechanisms, and coupled-mode analysis," Physical Review A **92**, 013836 (2015).

50. R. Van Laer, R. Baets, and D. Van Thourhout, "Unifying Brillouin scattering and cavity optomechanics," Physical Review A **93**, 053828 (2016).

51. P. Rakich, and F. Marquardt, "Quantum theory of continuum optomechanics," New Journal of Physics **20**, 045005 (2018).

52. R. C. Williamson, and R. D. Esman, "RF photonics," Journal of Lightwave Technology **26**, 1145-1153 (2008).

53. J. P. Yao, "Microwave Photonics," Journal of Lightwave Technology **27**, 314-335 (2009).

54. P. S. Devgan, D. P. Brown, and R. L. Nelson, "RF performance of single sideband modulation versus dual sideband modulation in a photonic link," Journal of Lightwave Technology **33**, 1888-1895 (2015).

55. D. Marpaung, J. Yao, and J. Capmany, "Integrated microwave photonics," Nature Photonics **13**, 80-90 (2019).

56. M. J. A. Schuetz, E. M. Kessler, G. Giedke, L. M. K. Vandersypen, M. D. Lukin, and J. I. Cirac, "Universal Quantum Transducers Based on Surface Acoustic Waves," Phys Rev X **5**, 031031 (2015).

57. R. Manenti, M. J. Peterer, A. Nersisyan, E. B. Magnusson, A. Patterson, and P. J. Leek, "Surface acoustic wave resonators in the quantum regime," Physical Review B **93** (2016).